\documentclass[preprint2]{aastex}
\usepackage[]{natbib}

\newcommand{\etal}{et al.}
\newcommand{\per}{\ensuremath{^{-1}}}
\newcommand{\persq}{\ensuremath{^{-2}}}

\newcommand{\percucm}{cm\ensuremath{^{-3}}}
\newcommand{\hal}{H\ensuremath{\alpha}}
\newcommand{\hbeta}{H\ensuremath{\beta}}
\newcommand{\hgamma}{H\ensuremath{\gamma}}
\newcommand{\hst}{\emph{HST}}
\newcommand{\msun}{\ensuremath{M_{\odot}}}
\newcommand{\zsun}{\ensuremath{Z_{\odot}}}
\newcommand{\kms}{km s\ensuremath{^{-1}}}

\newcommand{\lam}{\ensuremath{\lambda}}
\newcommand{\lamlam}{\ensuremath{\lambda\lambda}}

\newcommand{\starb}{\textsc{STARBURST99}}
\newcommand{\mup}{\ensuremath{M_{\rm up}}}
\newcommand{\teff}{\ensuremath{T_{\rm eff}}}
\newcommand{\nh}{\ensuremath{n_{\rm H}}}

\slugcomment{}
\shorttitle{}
\shortauthors{Barth \& Shields}

\begin{document}

\title{LINER/\ion{H}{2} ``Transition'' Nuclei and the Nature of NGC
4569}

\author{Aaron J. Barth}
\affil{Harvard-Smithsonian Center for Astrophysics, 60 Garden Street,
Cambridge, MA 02138}
\email{abarth@cfa.harvard.edu}

\author{Joseph C. Shields}
\affil{Department of Physics and Astronomy, Ohio University,
Clippinger Labs 251B, Athens OH 45701}
\email{shields@helios.phy.ohiou.edu}

\begin{abstract}

Motivated by the discovery of young, massive stars in the nuclei of
some LINER/\ion{H}{2} ``transition'' nuclei such as NGC 4569, we have
computed photoionization models to determine whether some of these
objects may be powered solely by young star clusters rather than by
accretion-powered active nuclei.  The models were calculated with the
photoionization code CLOUDY, using evolving starburst continua
generated by the the \starb\ code of \citet{lei99}.  We find that the
models are able to reproduce the emission-line spectra of transition
nuclei, but only for instantaneous bursts of solar or higher
metallicity, and only for ages of $\sim3-5$ Myr, the period when the
extreme-ultraviolet continuum is dominated by emission from Wolf-Rayet
stars.  For clusters younger than 3 Myr or older than 6 Myr, and for
models with a constant star-formation rate, the softer ionizing
continuum results in an emission spectrum more typical of \ion{H}{2}
regions.  This model predicts that Wolf-Rayet emission features should
appear in the spectra of transition nuclei.  While such features have
not generally been detected to date, they could be revealed in
observations having higher spatial resolution.  Demographic arguments
suggest that this starburst model may not apply to the majority of
transition nuclei, particularly those in early-type host galaxies, but
it could account for some members of the transition class in hosts of
type Sa and later.  The starburst models during the
Wolf-Rayet-dominated phase can also reproduce the narrow-line spectra
of some LINERs, but only under conditions of above-solar metallicity
and only if high-density gas is present ($n_e \gtrsim 10^5$ \percucm).
This scenario could be applicable to some ``Type 2'' LINERs which do
not show any clear signs of nonstellar activity.

\end{abstract}


\section{Introduction}

Emission-line nebulae in galactic nuclei are generally considered to
fall into three major categories: star-forming or \ion{H}{2} nuclei,
Seyfert nuclei, and low-ionization nuclear emission-line regions, or
LINERs.  The formal divisions between these classes are somewhat
arbitrary, as the observed emission-line ratios of nearby galactic
nuclei fall in a continuous distribution between LINERs and Seyfert
nuclei and between LINERs and \ion{H}{2} nuclei
\citep[e.g.,][]{hfs93}.  Traditionally, LINERs have been defined as
those nuclei having emission-line flux ratios which satisfy the
relations [\ion{O}{2}] \lam3727/[\ion{O}{3}] \lam5007 $> 1$ and
[\ion{O}{1}] \lam6300/[\ion{O}{3}] \lam5007 $> 1/3$ \citep{h80}.  It
is possible to construct alternative but practically equivalent
definitions, based on other line ratios, which can be applied to
datasets that do not include the wavelengths of the [\ion{O}{1}],
[\ion{O}{2}], or [\ion{O}{3}] lines \citep[e.g.,][]{hfs97a}.

A sizeable minority of galactic nuclei has emission-line ratios which
are intermediate between those of ``pure'' LINERs and those of typical
\ion{H}{2} regions powered by hot stars; these galaxies would be
classified as LINERs except that their [\ion{O}{1}] \lam6300 line
strengths are too small in comparison with other lines to meet the
formal LINER criteria.  Objects falling into this category have been
dubbed ``transition'' galaxies by \citet{hfs93}, and although this
nomenclature is somewhat ambiguous we adopt it here for consistency
with the spectroscopic survey of \citet{hfs97a}.  That survey defined
the transition class in terms of the following flux ratios:
\begin{center}
{[\ion{O}{3}]} \lam5007/\hbeta\ $<$ 3, \\
0.08 $\leq$ [\ion{O}{1}] \lam6300/\hal\ $<$ 0.17, \\ 
{[\ion{N}{2}]} \lam6583/\hal\ $\geq$ 0.6, \\
{[\ion{S}{2}]} \lamlam6716, 6731 /\hal\ $\geq$ 0.4.
\end{center}
\citet{ft92} have used the term ``weak-[\ion{O}{1}] LINERs'' to refer
to galaxies having [\ion{N}{2}] \lam6583/\hal\ $\gtrsim 0.6$ (typical
of LINERs) but which have [\ion{O}{1}] \lam6300/\hal\ $< 1/6$.  This
category is essentially identical to the transition class of
\citet{hfs93,hfs97a}, and we will refer to these galaxies as
transition objects in this paper.

According to the survey results of \citet{hfs97b}, this transition
class accounts for 13\% of all nearby galaxies, making them about as
numerous as Seyfert nuclei.  The Hubble type distribution of
transition galaxies is intermediate between that of LINERs, which are
most common in E/S0/Sa galaxies, and that of \ion{H}{2} nuclei, which
occur most often in Hubble types later than Sb \citep{hfs97b}. Roughly
$20\%$ of galaxies with Hubble types ranging from S0 to Sbc belong to
the transition class.  There is not a consensus, however, as to
whether these transition objects should be regarded as star-forming
nuclei, as accretion-powered active nuclei, or as composite objects
powered by an AGN and by hot stars in roughly equal proportion.
 
There is a large body of literature on the subject of the excitation
mechanism of LINERs which is relevant to the similar transition
class.  A variety of physical mechanisms has been proposed to explain
the emission spectra of LINERs, including shocks, photoionization by a
nonstellar ultraviolet (UV) and X-ray continuum, and photoionization
by hot stars.  (See Filippenko 1996 for a review.)  The possibility
that LINERs (and Seyfert nuclei as well) might be photoionized by
starlight was raised by \citet{tm85}, who suggested that very hot
($\teff \sim 10^5$ K) Wolf-Rayet (W-R) stars in a metal-rich starburst
could give rise to an ionizing continuum with a nearly power-law shape
in the extreme-UV.  More recent atmosphere models have indicated
substantially lower temperatures for W-R stars, however, casting doubt
on the Warmer hypothesis \citep{lgs92}.  Subsequent photoionization
models have attempted to explain LINER and transition-type spectra as
resulting from massive main-sequence stars.  \citet{ft92} found that
the spectra of weak-[\ion{O}{1}] LINERs could be explained in terms of
photoionization by O3--O4 stars having effective temperatures of
$\gtrsim45,000$ K, at ionization parameters of $U \approx 10^{-3.7}$
to $10^{-3.3}$.  \citet{shi92} carried this line of argument farther,
proposing that genuine LINER spectra could be generated by early O
stars with $\teff \approx 50,000$ K, provided that a high-density
component ($n_e \approx 10^{5.5}$ \percucm) is present in the NLR; the
high densities are needed to boost the strengths of high
critical-density emission lines, most notably [\ion{O}{1}]
\lam6300.  Similar conclusions were reached by \citet{sf94}, who
explored the effects of absorption by ionized gas as a means to harden
the effective ionizing spectrum.  Recent observations, particularly in
the UV and X-ray bands, have provided convincing evidence that many
LINERs are in fact AGNs, particularly the ``Type 1'' LINERs which have
a broad component to the \hal\ emission line \citep[for a recent
review see][]{ho99}.  The possibility has remained, however, that some
LINERs and transition nuclei are powered entirely by bursts of star
formation.  

An important shortcoming of the model calculations performed by
\citet{ft92} and \citet{shi92} is that the ionizing continua used as
input were those of single O-type stars; these studies did not address
the question of whether a LINER or transition-type spectrum could
result from the the \emph{integrated} ionizing continuum of a young
stellar cluster.  Compared with these single-star models, the
contribution of late-O and B stars will soften the ionizing spectrum,
making the emission-line ratios tend toward those of normal \ion{H}{2}
regions.  W-R stars, on the other hand, will harden the ionizing
spectrum during the period when these stars are present, roughly $3-6$
Myr after the burst.  Another drawback of the O-star models is that
they require the presence of stars with effective temperatures higher
than are thought to occur in \ion{H}{2} regions of solar or
above-solar metallicity, in order to produce a LINER or
transition-type spectrum rather than an \ion{H}{2} region spectrum.
Their applicability to galactic nuclei is therefore somewhat unclear.

Other mechanisms have been proposed for generating LINER or
transition-type spectra.  Shock excitation by supernova remnants in an
aging starburst may give rise to some transition objects; the nucleus
of NGC 253 is a likely candidate for such an object \citep{eng98}.
Also, post-AGB stars and planetary nebula nuclei will produce a
diffuse ionizing radiation field which could be responsible for the
very faint LINER emission (with \hal\ equivalent widths of $\sim1$
\AA) observed in some ellipticals and spiral bulges \citep{bin94}.

An alternate possibility is that the transition galaxies may simply be
composite systems consisting of an active nucleus surrounded by
star-forming regions.  For a galaxy at a distance of 10 Mpc, for
example, a 2\arcsec-wide spectroscopic aperture will include
\ion{H}{2} regions within 50 pc of the nucleus.  Galaxies having
emission lines both from a LINER nucleus and from surrounding
star-forming regions, in roughly equal proportions, will appear to
have a transition-type spectrum.  This interpretation was advocated by
\citet{hfs93} as the most likely explanation for the majority of
transition galaxies, and is consistent with the observed Hubble type
distribution for the transition class.  Other authors have similarly
contended that transition galaxies are AGN/\ion{H}{2} region
composites, based on optical line-profile decompositions
\citep{vgv97,gvv99} and near-infrared spectra \citep{hil99}.  Two of
the 65 transition nuclei observed in the \citet{hfs97b} survey have a
broad component to the \hal\ emission line, indicating the likely
presence of an AGN, and it is probable that many more transition
nuclei contain obscured AGNs which were not detected in the optical
spectra.  On the other hand, radio observations do not appear to
support the composite AGN/starburst interpretation.  In a VLA survey
of nearby galactic nuclei, Nagar \etal\ (1999) find compact,
flat-spectrum radio cores in more than 50\% of LINER nuclei, but in
only 6\% (1 of 18) of transition objects.  This discrepancy suggests
that the simple picture of an ordinary LINER surrounded by
star-forming regions may not apply to the majority of transition
objects.
  
Recent results from the \emph{Hubble Space Telescope} (\hst) have shed
new light on the question of the excitation mechanism of transition
nuclei.  As shown by \citet{mao98}, the UV spectrum of the well-known
transition nucleus in NGC 4569 over 1200-1600 \AA\ is virtually
identical to that of a W-R knot in the starburst galaxy NGC 1741,
indicating that O stars with ages of a few Myr dominate the UV
continuum.  \citet{mao98} find that the nuclear star cluster in NGC
4569 is producing sufficient UV photons to ionize the surrounding
narrow-line region, a key conclusion which provides fresh motivation
to study stellar photoionization models.  The brightness of the NGC
4569 nucleus, and the consequently high S/N observations that have
been obtained, make it one of the best objects with which to study the
transition phenomenon.

The recent availability of the \starb\ model set \citep{lei99} has
prompted us to reexamine the issue of ionization by hot stars in
LINERs and transition nuclei.  These models give predictions for the
spectrum and luminosity of a young star cluster, for a range of values
of cluster age, metal abundance, and stellar initial mass function
(IMF) properties.  Using the photoionization code CLOUDY \citep{fer98}
in combination with the \starb\ model continua, we have calculated the
expected emission-line spectrum of an \ion{H}{2} region illuminated by
a young star cluster, to test the hypothesis that some LINERs and
transition nuclei may be powered by starlight.  Similar calculations
have been performed by \citet{sl96}, but for the physically distinct
case of metal-poor objects representing \ion{H}{2} galaxies.  Other
examples of photoionization calculations for \ion{H}{2} regions using
evolving starburst continua are presented by \citet{gd94},
\citet{gbd95}, and \cite{bkg99}.

\section{The Nucleus of NGC 4569}
\label{section4569} 

Before describing the photoionization modeling, we review the
properties of NGC 4569, as it is among the best-known examples of the
transition class.  NGC 4569 is a Virgo cluster spiral of type Sab,
with a heliocentric velocity of $-235$ \kms, and we assume a distance
of 16.8 Mpc for consistency with the catalog of \citet{hfs97a}.  Its
nucleus is remarkably bright for a non-Seyfert, and so compact in the
optical that \citet{hum36} suspected it to be a foreground Galactic
star.  It is also an unusually bright UV source, with the highest 2200
\AA\ luminosity of the LINERs and transition objects observed by
\citet{mao95} and \citet{bar98}.  \hst\ Faint Object Spectrograph
(FOS) spectra show that the UV continuum is dominated by massive
stars, with prominent P Cygni profiles of \ion{C}{4} \lam1549,
\ion{Si}{4} \lam1400, and \ion{N}{5} \lam1240 \citep{mao98}.  The UV
spectrum is nearly an exact match to the spectrum of one of the
starburst knots in the W-R galaxy NGC 1741, an object with a likely
age in the range 3--6 Myr \citep{clv96}.  The optical spectrum of the
NGC 4569 nucleus is dominated by the light of A-type supergiants,
providing additional evidence for recent star formation \citep{kee96}.

One key result of the \citet{mao98} study was the conclusion that
the nuclear starburst in NGC 4569 is producing sufficient numbers of
ionizing photons to power the narrow-line region, assuming that the
surrounding nebula is ionization-bounded, \emph{even without
correcting for the effects of internal extinction on the UV continuum
flux}.  In fact, there appears to be substantial extinction within NGC
4569, as demonstrated by the UV continuum slope as well as the
presence of deep interstellar absorption features
\citep{mao98}. \citet{hfs97a} derive an internal reddening of $E(B-V)
= 0.46$ mag from the \hal/\hbeta\ ratio, while \citet{mao98} estimate
a UV extinction of $A \approx 4.8$ mag at 1300 \AA\ by comparison of
the observed UV slope with the expected spectral shape of an
unreddened starburst.

Despite the fact that NGC 4569 is often referred to as a LINER, and in
some cases presumed to contain an AGN on the basis of that
classification, there is no single piece of evidence which
conclusively demonstrates that an AGN is in fact present at all.  The
\hst\ images and spectra are all consistent with the nucleus being a
young, luminous, and compact starburst region.  No broad-line
component is detected on the \hal\ emission line \citep{hfs97a}, and
no narrow or broad emission lines are visible at all in the UV
spectrum other than the P Cygni features that are generated in O-star
winds \citep{mao98}.  Only the optical emission-line ratios point to a
possible AGN classification.  In a thorough study of optical and
\emph{IUE} UV spectra, \citet{kee96} concluded that there was at best
weak evidence for the presence of an AGN in NGC 4569, and that any AGN
continuum component, if present, must have an unusually steep spectrum.

Furthermore, while the nucleus of NGC 4569 is certainly extremely
compact, the UV and optical \hst\ images show that the nucleus is not
dominated by a central point source.  At 2200 \AA, the nucleus appears
extended in WFPC2 images with FWHM sizes of 13 and 9 pc along its
major and minor axes \citep{bar98}.  Optical WFPC2 images have been
discussed recently by \citet{pog99}, who state that the nucleus is
unresolved by \hst.  We have obtained these same images from the \hst\
archive.  While the nucleus is certainly compact, we find that it is
clearly extended even at the smallest radii.  A 12-second, CR-SPLIT
exposure in the F547M ($V$-band) filter is unsaturated and allows a
radial profile measurement.  We find a FWHM size of 14 pc by 8 pc
along the major and minor axes, consistent with the size of the
nuclear cluster measured at 2200 \AA.  \citet{bar98} estimated that at
most 23\% of the nuclear UV flux could come from a central point
source.  From the equivalent widths of stellar-wind features in the UV
spectrum, \citet{mao98} give a similar upper limit of $\sim20\%$ to
the possible contribution of a truly featureless continuum to the
observed UV flux.

X-ray observations of NGC 4569 with \emph{ROSAT} have revealed a
source coincident with the nucleus which is unresolved at the
2\arcsec\ resolution of the HRI camera \citep{cm99}.  This does not
necessarily indicate that an AGN is present, however, as the
optical/UV size of the starburst core is an order of magnitude smaller
than the HRI resolution.  \emph{ASCA} observations show that the X-ray
emission is extended over arcminute scales in both the hard (2--7 keV)
and soft (0.5--2 keV) bands \citep{ter99}.  Interestingly, the compact
source seen in the \emph{ROSAT} image is detected only in the soft
\emph{ASCA} band, while there is no detectable contribution from a
compact, hard X-ray source.  The spectral shape of the compact soft
X-ray component is consistent with an origin either in an AGN or in
X-ray binaries \citep{th99}, but the lack of a compact hard X-ray
source argues against the AGN interpretation.  If an AGN is present,
it must be highly obscured even at hard X-ray energies, with an
obscuring column of $N_H > 10^{23}$ cm\persq\ \citep{ter99}.  In radio
emission, VLA observations show that the NGC 4569 nucleus is an
extended source with a size of 4\arcsec\ and no apparent core
\citep{nh92}, in contrast with the compact, AGN-like cores found in
some LINERs \citep{fal98}.

Shock excitation has often been considered as a mechanism to power the
narrow emission lines in LINERs.  However, the lack of narrow emission
features in the UV spectrum of NGC 4569 argues against shock-heating
models for this object, as existing shock models generally predict
strong UV line emission \citep[e.g.,][]{dop96}.  Shock-excited
filaments in supernova remnants show strong emission in
high-excitation UV lines such as \ion{C}{4} \lam1549 and \ion{He}{2}
\lam1640 \citep[e.g.,][]{bla91, bla95} which are altogether absent
from the NGC 4569 spectrum.  Similarly, the shock-excited nuclear disk
of M87 \citep{dop97} has a high-excitation UV line spectrum which
bears no resemblance to the NGC 4569 spectrum.  From an analysis of
infrared spectra, \citet{ah99} proposed that the NGC 4569 nucleus is
powered by an 8--11 Myr-old starburst, by a combination of stellar
photoionization and shock heating from supernova remnants.  While this
hypothesis may be applicable to some LINERs and transition galaxies,
the UV spectrum of NGC 4569 shown by \citet{mao98} is inconsistent
with a burst of such advanced age, as the P Cygni features of
\ion{C}{4}, \ion{Si}{4}, and \ion{N}{5} would have disappeared from a
single-burst population after about 6 Myr.

The overall picture emerging from these observations is that the NGC
4569 nucleus is a compact, luminous, and young starburst.  The only
reason to invoke the presence of an AGN at all would be to explain the
higher strengths of the low-ionization forbidden lines in comparison
with values observed in normal \ion{H}{2} nuclei.  If it were indeed
possible for a young starburst to produce transition or LINER-type
emission lines in the surrounding gas, then there would be no reason
to consider AGN models for NGC 4569.

\section{Photoionization Calculations}
\label{sectioncalc}

\subsection{The Ionizing Continuum}

As discussed by \citet{ft92} and \citet{shi92}, the key ingredient
necessary for generating a LINER or transition-type emission-line
spectrum is an ionizing continuum which is harder than that produced
by typical clusters of OB stars.  A harder continuum will produce a
more extended partially-ionized zone in the surrounding \ion{H}{2}
region, boosting the strength of the low-ionization lines which are
typical of LINER spectra: [\ion{O}{1}] \lam6300, [\ion{O}{2}]
\lam3727, [\ion{N}{2}] \lam\lam6548,6583, and [\ion{S}{2}]
\lam\lam6716,6731.

To represent the ionizing continuum of a young starburst, we have
chosen the \starb\ model set; we refer the reader to \citet{lei99} for
the details of the methods used to construct these models.  Briefly,
the \starb\ code employs the Geneva stellar evolution models of
\citet{mey94}, with enhanced mass-loss rates, for high-mass stars.
Atmospheres are represented by the models compiled by \citet{lcb97}
and \citet{sch92}.  Figures 1--12 of \citet{lei99} display the
spectral energy distributions of the \starb\ model clusters for a
range of burst ages and for a variety of initial conditions.  From the
figures, some important trends are readily apparent.  During the first
2 Myr after an instantaneous burst, the continuum is dominated by the
hottest O stars, and there is essentially no emission below 228 \AA,
corresponding to the ionization energy of He$^{+}$.  The appearance of
W-R stars during the period 3--5 Myr after the burst results in a
dramatic change in the UV continuum, as these stars emit strongly in
the He$^{++}$ continuum below 228 \AA.  From 6 Myr onwards, the W-R
stars disappear and the UV continuum rapidly fades and softens as the
burst ages.  Only the models with an upper mass limit of \mup\ =
100\msun\ generate the hard, W-R-dominated UV continuum; the model
sequences with \mup\ = 30\msun\ do not generate significant numbers of
photons below 228 \AA\ for any ages because the progenitors of W-R
stars are not present in the initial burst.  Constant star-formation
rate models with \mup\ = 100\msun\ form W-R stars continuously after 3
Myr, but the overall shape of the UV continuum is softer than in the
instantaneous burst models, because of the continuous formation of
luminous O stars.

These results provide a useful starting point for the photoionization
calculations.  If it is possible for the \ion{H}{2} region surrounding
a young cluster to resemble a LINER or transition object, then this is
most likely to occur when the ionizing continuum is hardest, when W-R
stars are present during $t \approx 3-5$ Myr after a burst.  Very
massive stars (in the range 30--100 \msun\ or greater) must be present
in the burst or else the requisite W-R stars will not appear.  The
formation of W-R stars is enhanced at high metallicity, so the ability
to generate a LINER or transition-type spectrum may be a strong
function of metal abundance as well as age.  

\subsection{Model Grid}

To create grids of photoionization models, we fed the UV continua
generated by the \starb\ models into the photoionization code CLOUDY
\citep[version 90.04;][]{fer98}.  For each time step, a grid of models
was calculated by varying the nebular density and the ionization
parameter, which is defined as the ratio of ionizing photon density to
the gas density at the ionized face of a cloud.  Real LINERs and
transition nuclei are likely to contain clouds with a range of values
of density and ionization parameter, and more general models
incorporating density and ionization stratification can be constructed
as linear combinations of these simple single-zone models.  

All models were run with the following range of parameters: burst age
from 1 to 10 Myr at increments of 1 Myr, with log $U$ ranging from
$-2$ to $-4$ at increments of 0.5, and a constant density ranging from
log (\nh/\percucm) = 2 to 6 at increments of 1.  As a starting point,
we computed a grid for an instantaneous burst with an IMF having a
power-law slope of $-2.35$, \mup\ = 100 \msun, solar metallicity in
stars and gas, and a single plane-parallel slab of gas with no dust;
we will refer to this as model grid A.  The solar abundance set was
taken from \citet{ga89} and \citet{gn93}.  Other grids were computed
as variations on this basic parameter set, with the following
modifications made in different model runs: a constant star-formation
rate; metallicity 0.2, 0.4, or 2\zsun\ in both stars and gas; and
spherical geometry for the nebula.  To assess the effects of the
highest-mass stars, we also ran custom model grids, via the \starb\
web site, with $\mup = 70$ and 120 \msun.

The depletion of heavy elements onto grains can result in marked
changes to the emergent emission-line spectrum of an \ion{H}{2}
region, both by removing gas-phase coolants from the nebula and by
grain absorption of ionizing photons, which will modify the effective
shape of the ionizing continuum.  In metal-rich \ion{H}{2} regions,
these effects will tend to boost the strengths of the low-ionization
emission lines relative to the dust-free case \citep{sk95}.  To assess
the effects of dust in transition nuclei, we calculated additional
model grids which included dust grains with a Galactic ISM dust-to-gas
ratio along with the corresponding gas-phase depletions.  The dusty
models were all calculated using the solar abundance set for the
undepleted gas.  Dust grains were assumed to have the optical
properties of Galactic ISM grains, as described by \citet{mrn77},
\citet{dl84}, and \citet{mr91}.  From the CLOUDY output files, we
tabulated the strengths relative to \hbeta\ of the major emission
lines which are prominent in LINERs.

The calculations were performed under the assumption that the
\ion{H}{2} region is ionization bounded.  For this case, the outer
extension of the cloud was set to be the radius at which $T_e$ falls
to 4000 K, beyond which essentially no emission is generated in the
optical or UV lines.  As a test, we ran a grid of models with the
stopping temperature set to 1000 K, and we verified that the
emission-line ratios were essentially identical to the default case of
4000 K stopping temperature.  We also verified that the important
diagnostic line ratios differed by $\lesssim0.1$ dex between the
spherical and plane-parallel cases when all other input parameters
were unmodified, and all results discussed in this paper refer to the
plane-parallel models.  In the calculations, the longest timescales
for atomic species to reach equilibrium were of order $10^3$ years,
much shorter than the evolution timescale of the stellar cluster,
justifying the assumption that each time step of the cluster evolution
could be used independently to calculate the nebular conditions.
Table \ref{table1} gives a summary of the model parameters, for the
model grids which appear in the following discussion.

\placetable{table1}

\section{Discussion}
\label{sectiondiscuss}

\subsection{Model Results}

The model results are displayed in Figures
\ref{oratio}--\ref{ohimetal}.  To compare the model outputs with the
observed properties of a variety of galaxy types, we have used the
emission-line data compiled by \citet{hfs97a}.  This catalog has the
advantages of a homogeneous classification system, small measurement
aperture ($2\arcsec \times 4\arcsec$), and careful starlight
subtraction to ensure accurate emission-line data.  In order to reduce
confusion and to keep the sample of comparison objects to a reasonable
number, we included only objects with unambiguous classifications as
\ion{H}{2}, LINER, transition, or Seyfert.  Objects with borderline or
ambiguous classifications, such as ``LINER/Seyfert,'' were excluded
for clarity.  The comparison sample was further reduced by excluding
galaxies in which any of the emission lines \hal, \hbeta, [\ion{O}{3}]
\lam5007, [\ion{O}{1}] \lam6300, [\ion{N}{2}] \lam6583, or
[\ion{S}{2}] \lam\lam6716,6731 was undetected or was flagged as having
a large uncertainty in flux (``b'' or ``c'' quality flags).  The
measured line ratios are corrected for both Galactic and internal
reddening. 

Figure \ref{oratio} plots the ratio [\ion{O}{3}] \lam5007/\hbeta\
against [\ion{O}{1}] \lam6300/\hal\ at a burst age of 4 Myr, for the
solar-metallicity model grids A, B, C, and D.  The model results are
plotted for a density of $\nh = 10^3$ \percucm, as an approximate
match to the density of $n_e = 600$ \percucm\ measured for NGC 4569
\citep{hfs97a}.  The diagram shows that the instantaneous burst models
(A and B) are a good match to the line ratios of the transition
nuclei, for log $U \approx -3.5$.  The dusty models from grid B fall
more centrally within the region defined to contain transition
objects, but the models without dust still closely match transition
nuclei having lower [\ion{O}{1}] \lam6300/\hal\ ratios.  Figures
\ref{nratio} and \ref{sratio} show the corresponding diagrams for
[\ion{N}{2}] \lam6583/\hal\ and for [\ion{S}{2}] \lamlam6716,
6731/\hal, respectively.  In both cases we find that the single-burst
models span the region occupied by transition nuclei in the diagnostic
diagrams.

In the constant star-formation rate models (C and D), the UV continuum
remains softer than in the W-R-dominated phase of the instantaneous
burst models, because of the ongoing formation of luminous O stars.
As a result, the low-ionization emission lines are significantly
weaker than in the instantaneous burst models at 4 Myr.  These
constant star-formation rate sequences are a reasonable match to the
region of \ion{H}{2} nuclei in the diagram, and for [\ion{N}{2}]/\hal\
and [\ion{S}{2}]/\hal\ the agreement with \ion{H}{2} nuclei is
improved at lower densities of $n_e = 10^2$ \percucm, a value more
typical of \ion{H}{2} nuclei.  These constant star-formation rate
models are probably appropriate for galaxies having spatially
extended, ongoing star formation in their nuclei.

We note that the model curves shown in the figures should not be
expected to follow the locus of \ion{H}{2} nuclei in each plot,
despite the fact that the models are generated with a starburst
continuum.  The range of line ratios observed in \ion{H}{2} regions is
primarily a sequence in metal abundance \citep{mrs85}, while our
models are shown as sequences in $U$ for a given metallicity and
density.  Another point to note about the diagrams is that some of the
transition galaxies fall outside the region nominally defined for
transition objects, particularly in the [\ion{O}{1}]/\hal\ ratio
(Figure \ref{oratio}).  These galaxies were classified as transition
objects by \citet{hfs97a} on the basis of meeting the majority of the
classification criteria.  Similarly, some overlap can be seen in the
diagrams between the regions occupied by LINERs and Seyfert nuclei;
this again reflects the fact that galaxies span a continuous range in
the values of these emission-line ratios.

The variation of [\ion{O}{1}] line strength as a function of density
is shown in Figure \ref{odensity} for model grid A at $t = 4$ Myr.
For the range of densities considered ($\nh = 10^2$ to $10^6$
\percucm), the models closely overlap the transition region in the
diagram at log $U \approx -3.5$.  Introducing ISM depletion and dust
grains to the nebula primarily increases the [\ion{O}{1}]/\hal\ ratio
at low density.  As expected, the [\ion{O}{1}]/\hal\ ratio increases
with \nh\ up to densities of $10^5$ \percucm, while at densities
approaching the critical density of the \lam6300 transition ($1.6
\times 10^6$ \percucm) this ratio saturates and begins to turn over.

Figure \ref{oage} shows the [\ion{O}{1}]/\hal\ ratio as a function of
burst age (at $n_e = 10^3$ \percucm) for model grid A, for ages of 2
to 6 Myr.  This diagram highlights the dramatic changes that W-R stars
generate in the surrounding nebula.  From 3 to 5 Myr after the burst,
when W-R stars dominate the UV continuum, the harder ionizing
continuum boosts the strength of [\ion{O}{1}] by an order of magnitude
and the model sequences appear adjacent to the transition region, with
relatively little evolution in the line ratios during this period.  As
the burst ages beyond 6 Myr, the [\ion{O}{1}]/\hal\ ratio continues to
fall, and the emission-line strengths drop rapidly as the ionizing
continuum softens and its luminosity decreases.  At $Z = 2\zsun$, the
WR-dominated phase occurs slightly later, during time steps 4, 5, and
6 Myr.  The [\ion{N}{2}]/\hal\ and [\ion{S}{2}]/\hal\ ratios have a
similar dependence on burst age, and are displayed in Figures
\ref{nage} and \ref{sage}, respectively; these results are quite
similar to the calculations presented by \citet{lgs92} to illustrate
the effects of the W-R continuum on the [\ion{N}{2}]/\hal\ ratio.
While the [\ion{O}{1}]/\hal\ ratios in the models at $\log U = -3.5$
are too low by $\sim0.1-0.2$ dex to fit within the nominal transition
region, they still closely match those transition nuclei having
relatively low values for [\ion{O}{1}]/\hal, and the low-ionization
line strengths can be further enhanced by the inclusion of dust and
depletion (as in Figure \ref{oratio}).

One puzzling aspect of Figure \ref{oage} is that for ages outside the
range 3--5 Myr, the models predict [\ion{O}{1}]/\hal\ ratios too low
to match the majority of \ion{H}{2} nuclei.  \citet{sl96} and
\citet{m97} discuss this same problem in the context of
low-metallicity starburst galaxies.  They propose that shocks
generated by supernovae and stellar winds provide the additional
[\ion{O}{1}] emission, without making a significant contribution to
the [\ion{O}{2}] or [\ion{O}{3}] line strengths.  Shocks could play a
similar role in transition galaxies, as in the model of \cite{ah99},
although the lack of high-excitation UV line emission in transition
nuclei is problematic for the shock hypothesis.  A higher upper mass
cutoff alleviates this problem to some extent, at least for very young
bursts.  Increasing \mup\ to 120 \msun\ boosts [\ion{O}{1}]/\hal\ by
$\sim0.2$ dex for ages of $\lesssim3$ Myr.  Due to the very short
lifetimes of the highest-mass stars, however, the \mup\ = 120, 100,
and 70 \msun\ model grids result in identical emission-line spectra
from $\sim4$ Myr onward.

Metal abundance is an additional parameter which must be considered.
The models displayed up to this point were all calculated for a solar
abundance set, while the nuclei of early-type spirals are likely to
have enhanced heavy-element abundances.  As discussed by
\citet{lei99}, the continuum shortwards of 228 \AA\ is strongest in
the high-metallicity models, because the increased mass-loss rates
lead preferentially to the formation of W-R stars at high metal
abundance.  Figure \ref{olometal} plots the [\ion{O}{1}]/\hal\ ratio
for abundances of $Z$ = 0.2, 0.4, 1, and 2 \zsun, at a density of $\nh
= 10^3$ \percucm\ and $t = 4$ Myr.  From this diagram, it is clear
that solar or higher abundances are necessary to match the
[\ion{O}{1}] strengths of the transition nuclei; at lower abundances
the line ratios are a better match to those of the high-excitation
(low-metallicity) \ion{H}{2} nuclei.  Figure \ref{ohimetal} displays
the density dependence of the [\ion{O}{1}]/\hal\ ratio for the $Z =
2\zsun$ model grid; by comparison with the solar-metallicity model
grid in Figure \ref{odensity}, the higher abundances result in a
lower-excitation spectrum with enhanced [\ion{O}{1}] emission, due to
the harder extreme-UV continuum.

It would be advantageous to compare the model results with a wider
variety of emission lines.  Unfortunately, measurements of other
optical emission lines are scarce for transition nuclei.  The Ho
\etal\ survey did not include the [\ion{O}{2}] \lam3727 line, and
there is no other homogeneous catalog of [\ion{O}{2}] measurements for
transition galaxies.  To be consistent with a LINER or transition-type
classification, a model calculation must result in the flux ratio
[\ion{O}{2}] \lam3727 / [\ion{O}{3}] \lam5007 $>1$.  In fact, all of
the models with log $U \leq -3$ do satisfy this criterion.  Thus, any
of our models which is consistent with the Ho \etal\ LINER or
transition classification criteria is also consistent with the
original \citet{h80} criterion for the [\ion{O}{2}]/[\ion{O}{3}] ratio
in LINERs.  The relative strengths of UV lines such as \ion{C}{2}]
\lam2326, \ion{C}{3}] \lam1909, and \ion{C}{4} \lam1549 can provide
further diagnostics, but none of these lines is detected in NGC 4569
\citep{mao98}.  The only other transition nucleus having \hst\ UV
spectra available is NGC 5055, and its spectrum appears to be devoid
of UV emission lines as well \citep{mao98}.

We ran one additional model grid to test whether different model
atmospheres for O stars would lead to different results.  The \starb\
continua were calculated using stellar atmosphere models compiled by
\citet{lcb97}, which are based on the \citet{kur92} model set for the
massive stellar component.  The recent CoStar model grid of
\citet{sdk97}, which includes non-LTE effects, stellar winds, and line
blanketing for O stars, makes dramatically different predictions for
the ionizing spectra.  As shown by \citet{sv98}, the CoStar models
yield a luminosity in the He$^{++}$ continuum which is four orders of
magnitude greater than that predicted by the Kurucz models, for the
most massive O stars which dominate the UV luminosity at burst ages of
$<3$ Myr.  In the CoStar-based models the photon output of the cluster
below 228 \AA\ is essentially constant from 0 to 5 Myr.  To
investigate the effects of this harder O-star continuum on the
emission-line spectra, we ran a grid of models using the evolving
starburst continua computed by \citet{sv98} with the CoStar
atmospheres.  Model parameters were the same as for model grid A
except that an upper mass limit of 120 \msun\ was used.  We find that
using the CoStar atmospheres has a relatively minor effect on our
results.  In comparison with the \starb-based models having \mup\ =
120\msun, the CoStar model grid yields an increase in the
[\ion{O}{1}]/\hal\ ratio of $\sim0.1-0.15$ dex for $t < 6$ Myr, while
[\ion{N}{2}]/\hal\ and [\ion{S}{2}]/\hal\ are essentially unaffected.
During the period $t < 3$ Myr, the CoStar-based models result in an
\ion{H}{2} region spectrum, demonstrating that W-R stars are still
required in order to generate LINER or transition-type line ratios.

The strength of the [\ion{Ca}{2}] emission lines at 7291 and 7324 \AA\
is often used as a diagnostic of dust and depletion, because in the
absence of depletion these lines are predicted to be strong in
photoionized gas \citep[e.g.,][]{kff95, vmb96}.  (The \lam7291 line is
a cleaner diagnostic since \lam7324 is blended with [\ion{O}{2}]
\lam7325.)  However, for a 4 Myr-old burst with nebular conditions of
$n_e = 10^3$ \percucm, $\log U = -3.5$, and an undepleted solar
abundance set, our calculations yield a maximum prediction of only 0.2
for the ratio of [\ion{Ca}{2}] \lam\lam7291, 7324 to \hbeta.  Only at
very low ionization parameters ($\lesssim 10^{-4.5}$) does the
[\ion{Ca}{2}] emission become stronger than \hbeta.  Since \hbeta\ is
only barely visible in the spectra of many transition objects (prior
to careful starlight subtraction, at least), typical observations may
not have sufficient sensitivity to detect faint [\ion{Ca}{2}] lines in
these objects.  High-quality spectra of LINERs do not show
[\ion{Ca}{2}] emission \citep{hfs93}, indicating that Ca is likely to
be depleted onto grains in these objects, but similar data are not
generally available for transition nuclei.  If the [\ion{Ca}{2}] lines
are found to indicate a high level of depletion onto dust grains in
transition nuclei, this would also provide a further argument against
shock-heating models, as shocks will tend to destroy grains
\citep[e.g.,][]{mrw96}.

\subsection{The Nature of Transition Nuclei}
\label{sectiontransition}

The results shown in the preceding figures demonstrate that the
starburst models are in fact able to reproduce the major diagnostic
emission-line ratios of transition nuclei with reasonable accuracy,
during the period $t$ = 3--5 Myr when W-R stars are present.  For a
density of $10^3$ \percucm\ and an age of 4 Myr, the solar-metallicity
models with and without depletion bracket the range of values observed
in real transition nuclei for the line ratios [\ion{O}{1}]/\hal,
[\ion{N}{2}]/\hal, and [\ion{S}{2}]/\hal.  We do not attempt to
fine-tune a model to produce an exact match with the spectrum of NGC
4569, but the basic solar-metallicity dust-free model at $t = 4$ Myr
with $n_{\rm H} = 10^3$ \percucm\ and log $U = -3.5$ closely fits the
observed [\ion{O}{1}]/\hal\ ratio, while overpredicting
[\ion{S}{2}]/\hal\ by $\sim0.2$ dex and underpredicting
[\ion{N}{2}]/\hal\ by $\sim0.1$ dex.

We emphasize that the starburst models are only able to produce
transition-type spectra for the case of an instantaneous burst; that
is, when the burst duration is shorter than the timescale for
evolution of the most massive stars.  Multiple-burst populations can
only yield a transition spectrum if the dominant population is
$\sim3-5$ Myr old and the older or younger bursts do not contribute
significantly to the ionizing photon budget.  Models with a constant
star-formation rate produce \ion{H}{2} region spectra at all ages, as
the softer ionizing continua do not produce sufficient [\ion{O}{1}]
\lam6300 emission in the surrounding \ion{H}{2} region to match
transition-type spectra.  The parameter which is most important for
determining the hardness of the ionizing continuum is the number ratio
of W-R stars to O stars, which exceeds $\sim0.15$ during the
W-R-dominated phase in the \starb\ models at solar metallicity, and
approaches or exceeds unity at $Z = 2\zsun$.  In the constant
star-formation rate models at solar metallicity, the W-R/O ratio
levels off at $\sim0.06$ after about 4 Myr.  The compact size of the
NGC 4569 nucleus is consistent with the requirement that the burst
duration must be brief ($\lesssim1$ Myr) in order to generate a
transition-type spectrum.  The FWHM size of the starburst core in NGC
4569 is only $\sim10$ pc.  For such a burst to occur in $\lesssim1$
Myr would require a propagation speed for star formation of only
$\sim10$ \kms.  In fact, the typical velocities in the NGC 4569
nucleus are much greater than 10 \kms: the [\ion{N}{2}] \lam6583 line
has a velocity width of 340 \kms\ \citep{hfs97a}.  Thus, the NGC 4569
nucleus could represent the result of a single, rapid burst of star
formation.

Although our results suggest that transition galaxy spectra may be
attributed to a starburst with a high W-R/O-star ratio, the
demographics of transition nuclei and \ion{H}{2} nuclei indicate that
many transition galaxies are probably not formed by this mechanism.
In the \starb\ models, the W-R-dominated phase in an instantaneous
burst lasts for $\sim3$ Myr (i.e., 3 time steps in the calculations).
An \ion{H}{2} region surrounding an instantaneous burst will be
visible for $\sim6$ Myr, after which the emission lines will fade
rapidly \citep[e.g.,][]{gd94}.  Thus, for an instantaneous burst
population, the transition phase and the \ion{H}{2} nucleus phase will
have approximately equal lifetimes.  If all \ion{H}{2} nuclei
consisted of instantaneous burst stellar populations with nebular
conditions conducive to the formation of transition-type spectra, then
\ion{H}{2} nuclei and transition nuclei should be roughly equal in
number.  In reality, it is likely that a large fraction of
star-forming nuclei contain multiple bursts of star formation and/or
conditions of low density or low metallicity, so all star-forming
nuclei should not be expected to evolve through a transition-type
phase.  Although it is difficult to make specific predictions, it is
probably safe to conclude that for a given Hubble type, transition
nuclei generated solely by starbursts should be considerably less
numerous than ordinary \ion{H}{2} nuclei.

The statistics compiled by \citet{hfs97b} provide a basis for
comparison.  In early-type galaxies (E and S0), transition nuclei
outnumber \ion{H}{2} nuclei by a 3-to-1 margin.  Only for Hubble types
Sb and later do \ion{H}{2} nuclei begin to outnumber transition nuclei
by a factor of 2 or more.  The most straightforward interpretation of
this trend is that in early-type host galaxies, the majority of
transition nuclei are actually AGN/\ion{H}{2} region composites, as
proposed by \citet{hfs93} and others.  At intermediate and late Hubble
types, the population of transition nuclei may consist of both
composite objects and ``pure'' starbursts evolving through the
W-R-dominated phase.

The presence of transition nuclei in a small fraction ($\sim10\%$) of
elliptical galaxies \citep{hfs97b} presents a particularly intriguing
problem.  The Ho \etal\ survey detected five transition nuclei in
ellipticals but not a single case of an elliptical galaxy hosting an
\ion{H}{2} nucleus.  Given that the models which have been considered
for transition nuclei involve star formation, either alone or in
combination with an AGN, this observation is rather puzzling.  Perhaps
faint AGNs in elliptical nuclei can produce transition-type spectra
without substantial star formation activity.  Four of the five
transition nuclei found in ellipticals by \cite{hfs97a} have
borderline or ambiguous spectroscopic classifications, however, so
``pure'' transition nuclei in ellipticals are evidently quite rare.

Given these results, one might expect to see transition-type emission
spectra in some fraction of disk \ion{H}{2} regions in spiral
galaxies, but in fact such spectra are never found.  Single-burst
models for disk \ion{H}{2} region spectra are only compatible with
observed line ratios for model ages of $t < 3$ Myr \citep{bkg99}, as
the harder ionizing spectrum after 3 Myr makes the models overpredict
the strengths of the low-ionization lines.  Bresolin \etal\ suggest
that either current stellar evolution models are at fault, or that
disk \ion{H}{2} regions are disrupted before reaching an age of 3 Myr,
in which case the W-R phase would not be observed in the nebular gas.
An alternate (and perhaps more attractive) possibility is that the
majority of \ion{H}{2} nuclei, as well as disk \ion{H}{2} regions, are
better described by the models with constant star-formation rate, or
contain multiple bursts of star formation with an age spread of a few
Myr, which would result in a spectrum similar to the constant
star-formation rate models. 

For understanding the physical nature of transition objects, the
observational challenge is to search for any unambiguous signs of
nonstellar activity.  Detection of broad \hal\ emission, or a compact
source of hard X-ray emission with a power-law spectrum, would provide
evidence for an AGN component.  High-resolution optical spectra (from
\hst) could provide a means to spatially resolve a central
AGN-dominated narrow-line region from the surrounding
starburst-dominated component.  Since direct evidence for
accretion-powered nuclear activity in transition nuclei is generally
lacking, it should not be assumed that any given transition object
actually contains an AGN unless observations specifically support that
interpretation.

One further effect that should be considered in starburst models in
the future is photoionization by the X-rays generated by the
starburst.  X-ray binaries and supernova remnants will provide
high-energy ionizing photons, resulting in a spatially extended source
of soft X-ray emission as observed in the nucleus of NGC 4569
\citep{ter99}, for example.  (The massive main sequence stars will
contribute only a negligible amount to the total X-ray luminosity of a
starburst; see Helfand \& Moran 1999.) Photoionization by X-rays will
naturally lead to an enhancement of the low-ionization forbidden
lines, and this could contribute to the excitation of some transition
galaxies.

\subsection{LINERs}
\label{sectionliner}

The strength of [\ion{O}{1}] \lam6300 is the key distinguishing factor
between LINERs and transition nuclei, and matching the observed
strength of this line is the major challenge of starburst models for
LINERs.  Our calculations show that LINER spectra can only be
generated by the \starb\ clusters under a very specific and limited
range of circumstances.  Model grids A and B, while matching the
[\ion{O}{1}] / \hal\ ratio of transition nuclei quite well, do not
overlap at all with the main cluster of LINERs in Figure \ref{oratio},
even at high densities and even when depletion and dust grains are
included.  Only grid G with $Z = 2\zsun$ is able to replicate the high
[\ion{O}{1}] / \hal\ ratios of most LINERs, and only during $t
\approx$ 4--6 Myr and at densities of \nh\ $\gtrsim 10^5$ \percucm.
In agreement with previous models, we find that values of log $U
\approx$ $-3.5$ to $-3.8$ reproduce the observed [\ion{O}{1}] / \hal\
ratios of LINERs.  However, at such high densities the models
underpredict the strengths of [\ion{S}{2}] and [\ion{N}{2}] relative
to \hal.  Single-zone models require $n_e \lesssim 10^5$ \percucm\ to
match the [\ion{N}{2}]/\hal\ ratios of LINERs and $n_e \lesssim 10^4$
\percucm\ for [\ion{S}{2}]/\hal.

Agreement with LINER spectra can be achieved with a simple two-zone
model, in which high-density and low-density components are present,
similar to the scenario proposed by \citet{shi92}.  As an example, a
two-component model constructed from grid A containing gas at ($n_e =
10^3$ \percucm, $U = 10^{-3.5}$) and at ($n_e = 10^5$ \percucm, $U =
10^{-4}$) produces emission-line ratios which are consistent with all
the LINER classification criteria of both the \citet{h80} and
\citet{hfs97a} systems, if the two density componenets are scaled so
as to contribute equally to the total \hbeta\ luminosity.  As a local
comparison, observations of near-infrared \ion{Fe}{2} emission
indicate the presence of clouds having $n_e > 10^5$ \percucm\ in the
Galactic center region \citep{dep92}, so it is plausible that other
galactic nuclei may contain ionized gas at similarly high densities
even in the absence of an observable AGN.  A starburst origin for some
LINER 2 nuclei would provide a natural explanation for the lower
values of the X-ray/\hal\ flux ratio seen in these objects, in
comparison with AGN-like LINER 1 nuclei \citep{ter99}.

It seems unlikely, however, that many LINERs are generated by this
starburst mechanism.  About 15\% of LINERs are known to have a broad
component of the \hal\ emission line, indicating a probable AGN
\citep{hfs97b}.  By analogy with the Seyfert population, a much larger
fraction of LINERs is likely to have broad-line regions which are
either obscured along our line of sight, or are simply too faint to be
detected in ground-based spectra against a bright background of
starlight.  Many LINERs show signs of nuclear activity that cannot be
explained by stellar processes: compact flat-spectrum radio sources or
jets, compact X-ray sources with hard power-law spectra, or
double-peaked broad Balmer-line emission, for example.  As an
increasing body of observational work supports the idea that many
LINERs are in fact AGNs, there is less incentive to consider purely
stellar models for their excitation.

Demographic arguments, similar to those given above for transition
nuclei, can be applied for the LINER population.  Since the LINER
phase only occurs for instantaneous bursts at high density and high
metallicity, the starburst scenario implies that \ion{H}{2} nuclei
should be considerably more numerous than starburst-generated LINERs
for a given Hubble type.  While LINERs are common in early-type hosts,
\ion{H}{2} nuclei are not found in elliptical hosts and are seen in
fewer than 10\% of S0 galaxies \citep{hfs97b}.  This disparity is a
strong argument against a starburst origin for those LINERs in
early-type galaxies.  In later Hubble types the situation is less
clear, however.  \ion{H}{2} nuclei occur in $\sim80\%$ of spirals of
type Sc and later, while LINERs occur in just 5\% of these galaxy
types \citep{hfs97b}.  It is conceivable that some of the LINERs in
intermediate to late-type hosts could have a starburst origin, and
this issue could be resolved by further UV and X-ray observations in
the future.  Interestingly, the Ho \etal\ survey did not find any
examples of broad \hal\ emission in LINERs or transition objects with
hosts of type Sc or later; perhaps star formation plays a more
prominent role than accretion-powered activity in these objects.  

While a few LINERs show spectral features of young stars in the UV
\citep{mao98}, the quality of the observational data is poor in
comparison with the NGC 4569 UV spectrum, and it is difficult to set
meaningful constraints on the age of the young stellar population.
NGC 404 is a possible candidate for a starburst-generated LINER, but
in its UV spectrum the P Cygni features are weak in comparison with
NGC 4569, indicating either an older burst population or dilution by a
featureless AGN continuum \citep{mao98}.  The LINERs having UV
spectral features from massive stars may also host obscured AGNs which
can be detected in other wavebands.  For example, the UV continuum of
the LINER NGC 6500 appears to have its origin in hot stars
\citep{bar97, mao98}, but observations of a parsec-scale radio jet
unambiguously demonstrate that nonstellar activity is occurring as
well \citep{fal98}.

\subsection{W-R Galaxies with LINER or Transition-Type Spectra}

The starburst models presented here run into two obvious problems.
First, W-R galaxies are almost never known to have LINER or
transition-type spectra.  Second, LINERs and transition nuclei almost
never show W-R features in their spectra.  Is there any way to
reconcile the starburst models with these facts?

W-R galaxies are identified by the appearance of the 4650 \AA\ blend
in their spectra \citep[e.g.,][]{ks81}.  Since the formation of W-R
stars is enhanced at high $Z$, the strength of this feature relative
to \hbeta\ increases dramatically with metallicity, from $\lesssim0.1$
at $Z < 0.4\zsun$ to $\sim0.5 - 4$ at $Z \geq \zsun$ \citep{sv98}.
However, in the nuclei of early-type spirals where high metallicities
are expected, a nuclear starburst will be surrounded by the old
stellar population of the galactic bulge, making the detection of the
W-R bump extremely difficult \citep{mkc99}.  Most of the currently
known W-R galaxies are late-type spirals or irregular galaxies
\citep{scp99} in which the W-R bump is visible against the nearly
featureless starburst continuum.  When the W-R bump is detected in
\ion{H}{2} galaxies, its amplitude above the continuum level is
generally far smaller than that of \hbeta\ or even \hgamma\
\citep[e.g.,][]{ks81}.  In most of the LINER and transition galaxy
spectra in the catalog of \citet{hfs95}, however, \hbeta\ barely
appears and \hgamma\ is too weak to be visible at all prior to
continuum subtraction.  Even in a high-metallicity environment, where
the total intensity of the W-R bump can be comparable to that of
\hbeta, the amplitude of the W-R bump above the continuum will be much
lower than that of \hbeta\ because the flux in the W-R feature is
spread over $\sim70-100$ \AA.  Thus, the detection of W-R emission in
galactic nuclei is strongly biased toward late-type, bulgeless
galaxies.  In late-type or dwarf irregular galaxies where the W-R bump
is visible, the W-R/O-star ratio is expected to be much smaller owing
to the lower metallicity, and the resulting softer ionizing spectrum
will tend to produce an \ion{H}{2} region spectrum rather than a
transition object.  The gas density as a function of Hubble type may
play a role as well; in a study of \ion{H}{2} nuclei, \citet{hfs97c}
find a weak trend toward lower nebular densities in later-type host
galaxies.

Observational detection of W-R features in transition nuclei is
perhaps the clearest test of the starburst models, if sufficiently
sensitive observations can be obtained.  The UV spectrum of NGC 4569
is consistent with an age of $\sim3-6$ Myr, an age at which W-R stars
are expected to be present.  Previous optical spectra have not
revealed the 4650 \AA\ W-R bump in NGC 4569, but further observations
with high S/N and small apertures would be worthwhile.  The lack of
\ion{He}{2} \lam1640 emission in the UV spectrum of NGC 4569 is
potentially a more serious problem since the burst population should
dominate at short wavelengths.  The models of \citet{sv98} predict an
equivalent width of at least 2 \AA\ in the W-R-generated \lam1640 line
during the period 3--6 Myr for an instantaneous burst of solar or
higher metallicity, while the observed upper limit of $f(1640) < 2.0
\times 10^{-15}$ erg s\per\ cm\persq\ \citep{mao98} corresponds to an
equivalent width limit of $\lesssim 0.3$ \AA.  It should be noted,
however, that the \lam1640 line lies at the extremely noisy blue end
of the FOS G190H grating setting, in a region where detection of
emission or absorption features is difficult.


Two W-R galaxies may provide useful points of reference for the
starburst models.  NGC 3367 is classified by \citet{hfs97a} as an
\ion{H}{2} nucleus on the basis of its [\ion{O}{1}]/\hal\ and
[\ion{S}{2}]/\hal\ ratios, although its [\ion{N}{2}]/\hal\ ratio of
0.83 is more consistent with a LINER or transition-type classification
and its emission lines are markedly broader than those of typical
\ion{H}{2} nuclei. \citet{ah99} describe NGC 3367 as a
starburst-dominated transition object \citep[see also][]{dda88}.  The
4650 \AA\ W-R bump was noted by \citet{hfs95}, who also suggested a
LINER/\ion{H}{2} classification and a composite source of ionization.
As a borderline \ion{H}{2} nucleus/transition object with clear
evidence for W-R stars, this object deserves further study, to
determine whether there is indeed an AGN or whether the enhanced
low-ionization emission may be the result of ionization by the W-R
population.  The electron density of 835 \percucm\ measured from
the [\ion{S}{2}] doublet \citep{hfs97a} is also noteworthy, as this is
among the highest densities found for an \ion{H}{2} nucleus in the Ho
\etal\ survey.

Another intriguing object is the nucleus of NGC 6764, which has been
classified variously as a Seyfert, a LINER, and a starburst by
different authors \citep[see][]{gvv99}.  This galaxy exhibits
prominent emission in the 4650 \AA\ W-R blend \citep{oc82}.  A recent
study by \citet{eck96} demonstrates that the narrow emission lines are
consistent with a LINER classification, but there are no unambiguous
signs of nonstellar activity in the nucleus.  \citet{eck96} find that
the nucleus contains $\sim3600$ W-R stars, and that the overall
properties of the object are consistent with ionization by the
starburst alone, rather than by a starburst/AGN composite.  If this
conclusion is confirmed by further observations, NGC 6764 could be
considered the best candidate for a LINER photoionized by a starburst
during its W-R-dominated phase.  \citet{ah99} derived an age of 9--10
Myr for the starburst in NGC 6764 based on near-infrared emission-line
diagnostics, but this age is inconsistent with the presence of W-R
stars, at least for the case of an instantaneous burst.

Compared with typical LINERs and transition nuclei, conditions for
detection of W-R spectral features in these two objects are perhaps
more favorable.  Both host galaxies are of late Hubble types (SBc for
NGC 3367 and SBbc for NGC 6764) in comparison with the majority of
LINERs and transition nuclei, so the level of contamination by the
surrounding old stellar population is relatively low.  Furthermore
(and partly as a result of this), the emission-line equivalent widths
in these two nuclei are relatively high for LINERs or transition
nuclei.  If the emission-line spectra of NGC 3367 or NGC 6764 had been
superposed on a luminous early-type spiral bulge, the W-R emission
might never have been noticed.

The detection of W-R emission in a LINER does not automatically imply
a purely starburst origin for the emission lines, of course, since
starbursts and AGNs are often known to coexist.  Mrk 477 is a
well-known example of a Seyfert 2 galaxy having a large population of
W-R stars in its nucleus \citep{h97}.  In the context of the starburst
model, however, the crucial test is to search for additional examples
of LINERs or transition nuclei which exhibit a high ratio of W-R to O
stars but no signs of accretion-powered activity.

\subsection{Caveats and Limitations}

The most important limitation of these calculations comes from the
accuracy of the input continua.  The conclusion that the starburst
models are able to reproduce transition or LINER spectra under some
circumstances depends crucially on the presence of W-R stars to
provide a hard and luminous ionizing continuum.  Unfortunately, the
continuum shape and luminosity of W-R stars in the extreme-UV band are
quite uncertain, particularly in the He$^{++}$ continuum where stellar
winds have a dramatic effect.  Several up-to-date reviews of the
numerous difficulties involved in modeling W-R spectra can be found in
the volume edited by \cite{vdh99}.  The \starb\ model grid uses the
W-R atmospheres of \citet{sch92}, which are calculated for a pure
helium composition, but more recent atmosphere models are beginning to
include the effects of line-blanketing, as well as clumping and
departures from spherical symmetry.  As discussed by \citet{lei99},
the W-R/O-star ratio is also extremely model dependent, and may be
revised in future generations of models.  This would have a direct
effect on the strength of the extreme-UV continuum and consequences
for the nebular emission lines.  Furthermore, the \starb\ models
neglect binary evolution, although this is more likely to affect the
W-R/O ratio at low metallicity.  Since the results of the
transition-object models are highly dependent on the most uncertain
portion of the W-R spectrum, new photoionization calculations should
be computed to assess the impact of different W-R evolution and
atmosphere models in the future.

The shape of the IMF in starburst regions is a subject of some debate,
and the possible variation of the IMF with metallicity is of
particular importance for galactic nuclei, which are likely to have $Z
> \zsun$.  \citet{k74} and \citet{st76} suggested that \mup\ should be
lower in regions of higher metal abundance, but this issue has not
been settled definitively.  Star-count observations demonstrate that
the IMF slope and \mup\ do not appear to vary with metallicity
\citep{mjd95}, at least for $Z \leq \zsun$.  While nebular diagnostics
in \ion{H}{2} galaxies are generally consistent with a Salpeter IMF
with $\mup \approx 100$ \msun\ at subsolar metallicity
\citep[e.g.,][]{sl96}, at high metallicity the observational situation
is somewhat ambiguous. \citet{bkg99} find that the mean stellar
temperature in \ion{H}{2} regions decreases significantly with
increasing $Z$, and that the \ion{He}{1} \lam5876 / \hbeta\ ratios of
\ion{H}{2} regions at $Z \approx 2\zsun$ are more consistent with
\mup\ = 30 \msun\ than with \mup\ = 100 \msun.  Such a low value for
\mup\ would pose serious difficulties for any starburst models of
LINERs and transition nuclei, as the massive progenitors of W-R stars
would not be present.  Counterbalancing this trend, the strong tidal
forces, turbulence, and magnetic field strengths in galactic nuclei
may act to raise the Jeans mass and favor the formation of more
massive stars \citep{mor93}.  In the Galactic center, there are stars
with initial masses of $\sim100$ \msun\ \citep{kra95}, and one
Galactic center object (the Pistol star) may have $M_{\rm initial}$ as
high as 200--250 \msun\ \citep{fig98}.  Thus, the proposed trend
toward lower values of \mup\ at high metallicity in disk \ion{H}{2}
regions may not apply to galactic nuclei.  Detailed comparison of the
UV spectra of galaxies such as NGC 4569 with starburst population
synthesis models can provide useful constraints on the population of
high-mass stars in nuclear starbursts.

Despite these uncertainties, these photoionization models have a major
advantage compared with previous generations of W-R or O-star models
for LINERs and transition nuclei, in that the \starb\ models with
standard parameters are constructed to represent the actual stellar
populations in starbursts, to the best of current knowledge.  Previous
O-star models \citep{ft92, shi92} required the presence of
hypothetical, unusually hot stars in order to explain LINER or
transition spectra, and they did not address the evolution of the
young stellar population at all.  The starburst models presented here
provide a more plausible mechanism to generate a transition-type
spectrum, even if this model may apply only to a relatively small
fraction of the population of transition galaxies.

\section{Conclusions}
\label{sectionconclusions}

Our primary conclusion is that for standard starburst parameters and
for nebular conditions which may be typical of galactic nuclei, the
starburst models are able to reproduce the important diagnostic
emission-line ratios for LINER/\ion{H}{2} transition galaxies,
otherwise known as weak-[\ion{O}{1}] LINERs.  The key ingredient
needed to generate a transition-type spectrum is a UV continuum
dominated by W-R stars, a condition which occurs during $t =$ 3--5 Myr
after an instantaneous burst.  A transition-type emission spectrum may
thus be a phase in the evolution of some nuclear \ion{H}{2} regions in
which the ionizing continuum is generated by a single-burst stellar
population.  The models are also able to produce an [\ion{O}{1}] /
\hal\ ratio high enough to match LINER spectra, but only for
conditions of above-solar metallicity combined with the presence of
high-density ($\gtrsim10^5$ \percucm) clouds.  A sensitive search for
W-R spectral features in transition nuclei would provide a test of
this starburst scenario.  This model may apply only to a small
fraction of LINERs and transition nuclei; many LINERs and some
transition objects show clear signs of nonstellar activity, and the
starburst models may not apply at all to objects in early-type host
galaxies.  Further multiwavelength observations of transition nuclei
will be of great utility for determining what fraction of them contain
genuine active nuclei, and what fraction appear to be purely the
result of stellar phenomena.

\acknowledgments

Research by A.J.B. is supported by a postdoctoral fellowship from the
Harvard-Smithsonian Center for Astrophysics.  This research was also
supported financially by grant AR-07988.02-96A, awarded to J.C.S. by
STScI, which is operated by AURA for NASA under contract NAS5-26555.
This work would not have been possible without the excellent
software created and distributed by Gary Ferland and the Cloudy team,
and by Claus Leitherer and the \starb\ team.  We also thank Gary
Ferland for providing a helpful referee's report, Claus Leitherer for
additional helpful comments on the manuscript, and Daniel Schaerer for
supplying model starburst spectra in electronic form.

\clearpage


\begin{center}
\begin{deluxetable}{lccc}
\tablewidth{3.5in}
\tablecaption{Model Parameters\label{table1}}
\tablehead{\colhead{Model Grid} & \colhead{SF Law} & 
        \colhead{$Z$/\zsun} & \colhead{Dusty?}}
\startdata
A & I & 1.0 & N \\
B & I & 1.0 & Y \\
C & C & 1.0 & N \\
D & C & 1.0 & Y \\
E & I & 0.2 & N \\
F & I & 0.4 & N \\
G & I & 2.0 & N \\
\enddata
\tablecomments{All models listed above are calculated for \mup\ = 100
\msun, IMF power-law slope of $-2.35$, and for plane-parallel nebular 
geometry. The star-formation law
in Column 2 denotes I = instantaneous burst, and C = constant
star-formation rate.  The metallicity $Z$ refers to the stars
 and to the undepleted abundance of the nebular gas.}
\end{deluxetable}
\end{center}
 
\clearpage


\begin{center}
\textbf{Figure Captions}
\end{center}


\figcaption[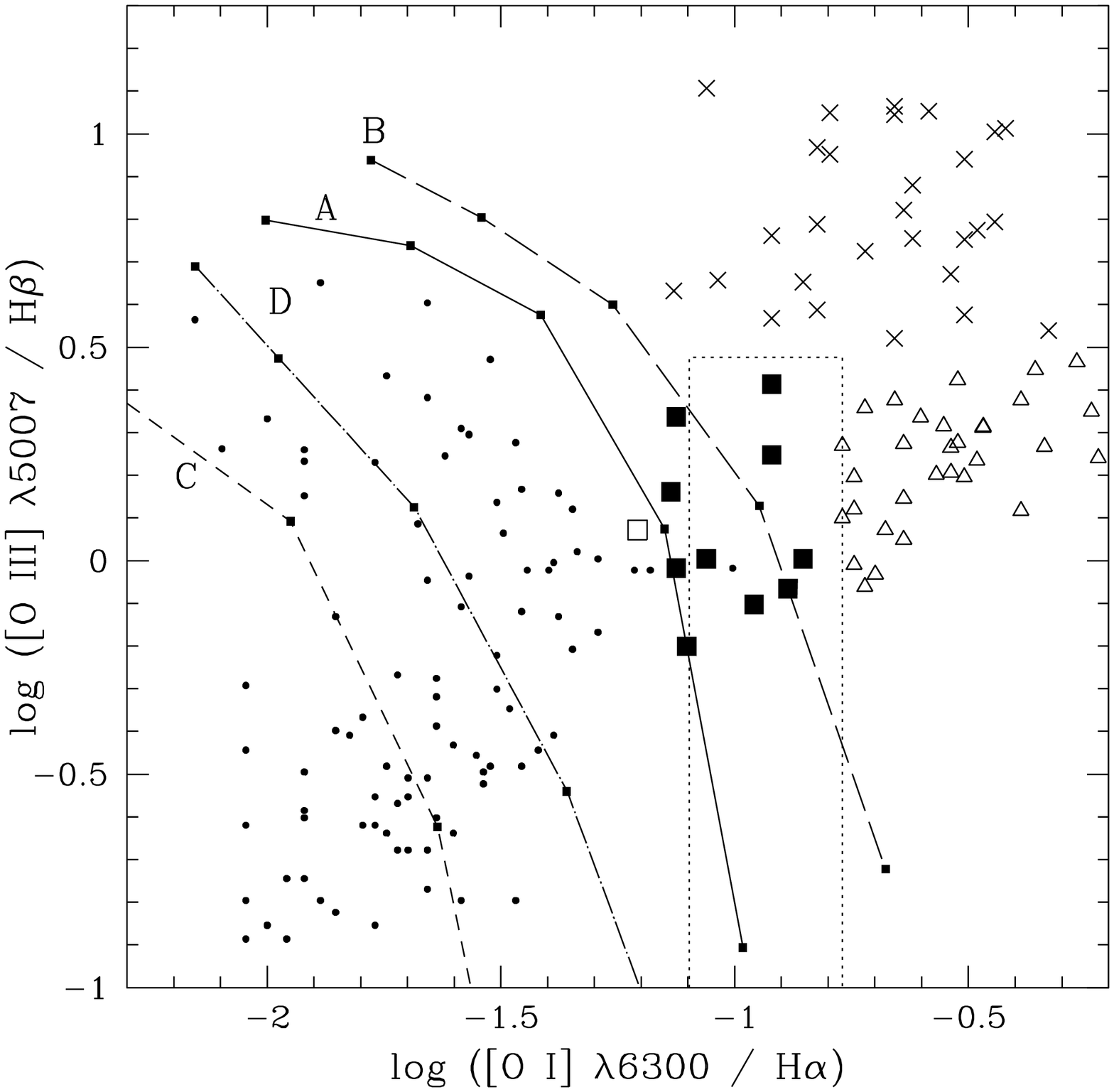]{\label{oratio} } \noindent Line-ratio diagram
of [\ion{O}{3}] \lam5007 / \hbeta\ against [\ion{O}{1}] \lam6300 /
\hal, for model grids A (solid line), B (long-dashed line), C
(short-dashed line), and D (dot-dashed line), at a burst age of 4 Myr
and $n_e = 10^3$ \percucm.  The input continuum has solar metallicity,
IMF power-law slope $-2.35$, and \mup\ = 100 \msun.  The following
description applies to this and all subsequent plots: the small
squares along each model line correspond to the model grid points at
log $U$ = $-4, -3.5, -3, -2.5$, and $-2$, with $U$ increasing upward
along the line.  The points plotted represent galaxies from the Ho
\etal\ catalog, as follows: \emph{Small circles}: \ion{H}{2} nuclei.
\emph{Squares:} LINER/\ion{H}{2} transition objects.  NGC 4569 is
represented by an open square.  \emph{Triangles:} ``pure'' LINERs.
\emph{Crosses:} Seyfert nuclei.  The dotted line encloses the region
defined for transition nuclei according to the criteria of
\citet{hfs97a}.

\figcaption[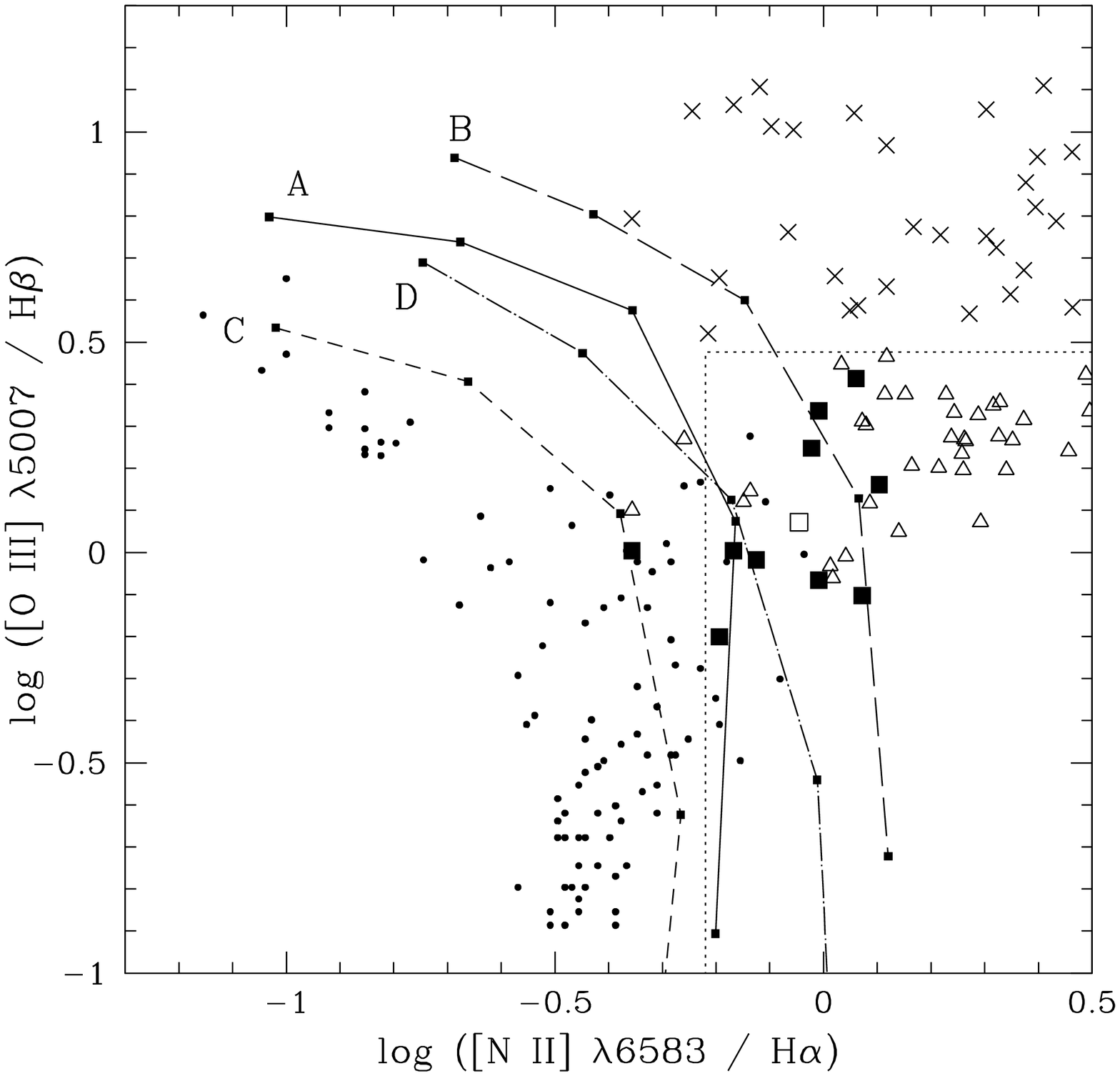]{Same as Figure \ref{oratio}, but for the line
ratio [\ion{N}{2}] \lam6583 / \hal.  The dotted line encloses the
region occupied by both LINERs and transition objects, according to
the criteria of \citet{hfs97a}.  \label{nratio}}  

\figcaption[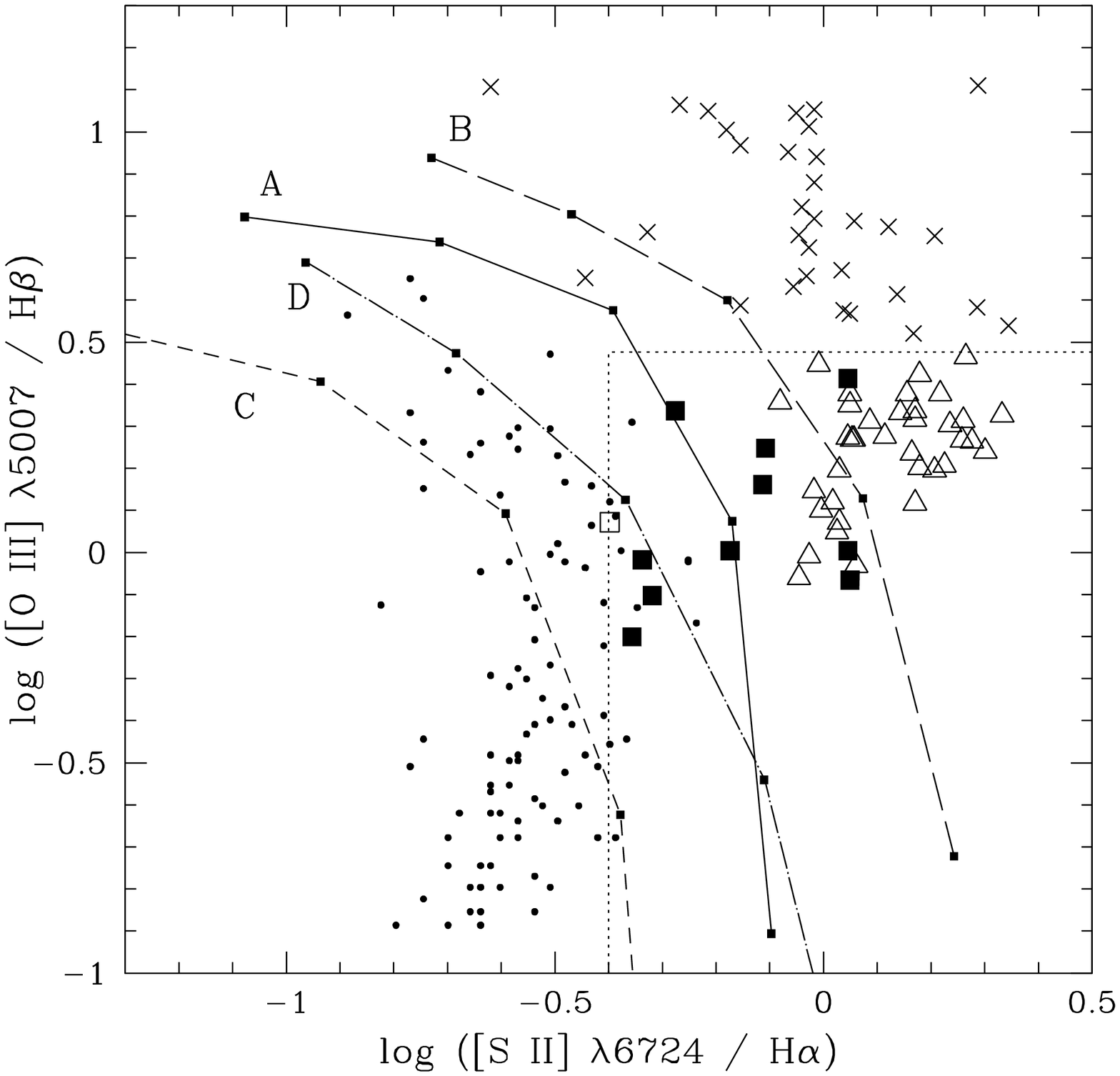]{Same as Figure \ref{oratio}, but for the line
ratio [\ion{S}{2}] \lamlam6716, 6731 / \hal.  The dotted line encloses
the region occupied by both LINERs and transition objects, according
to the criteria of \citet{hfs97a}.  \label{sratio}}

\figcaption[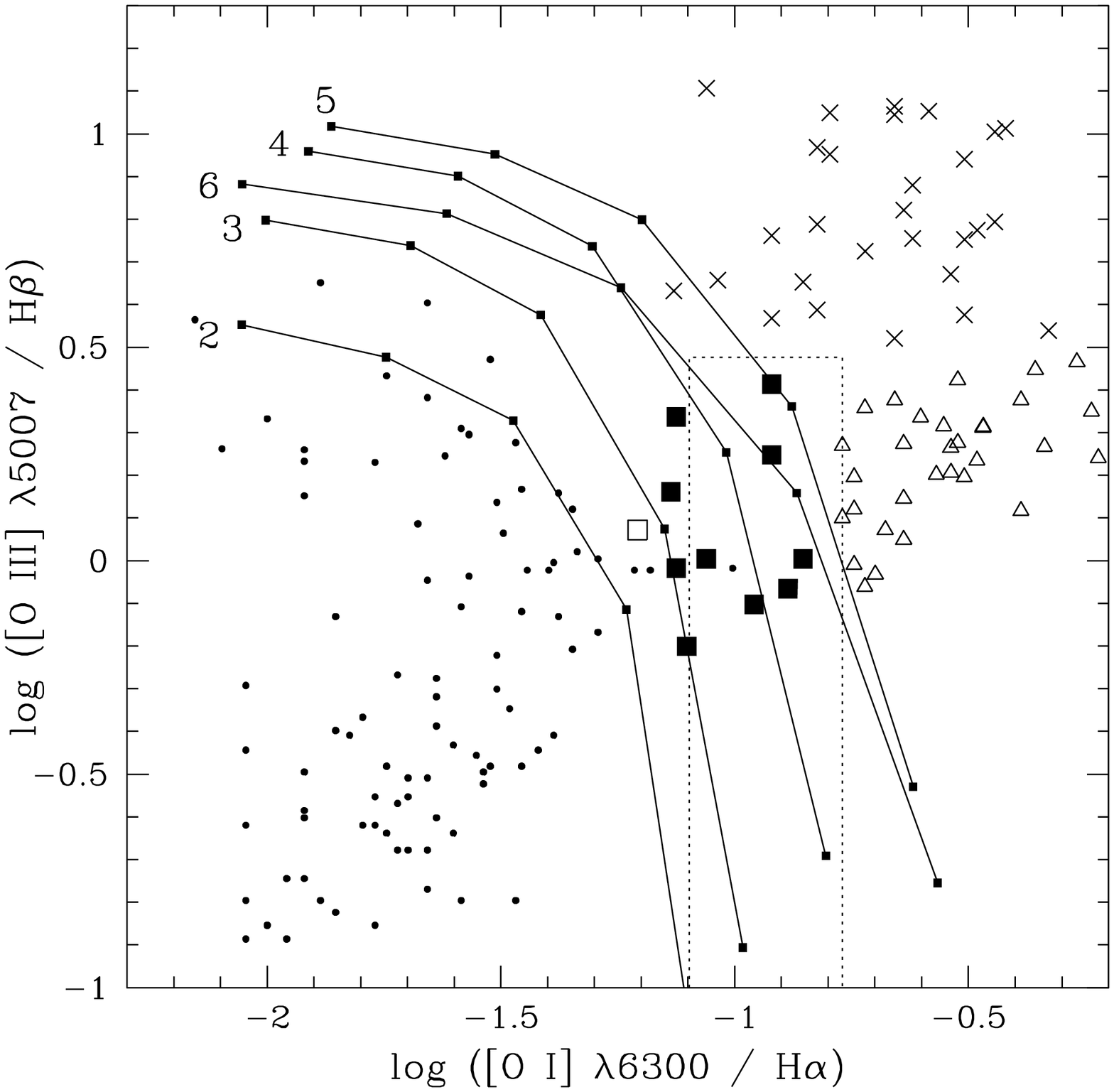]{Density effects on the [\ion{O}{1}] \lam6300
line strength, for model grid A.  Plot symbols for individual data
points are the same as for Figure \ref{oratio}.  The model sequences
are shown for age 4 Myr and nebular density of log (\nh/\percucm) = 2,
3, 4, 5, and 6. \label{odensity}}

\figcaption[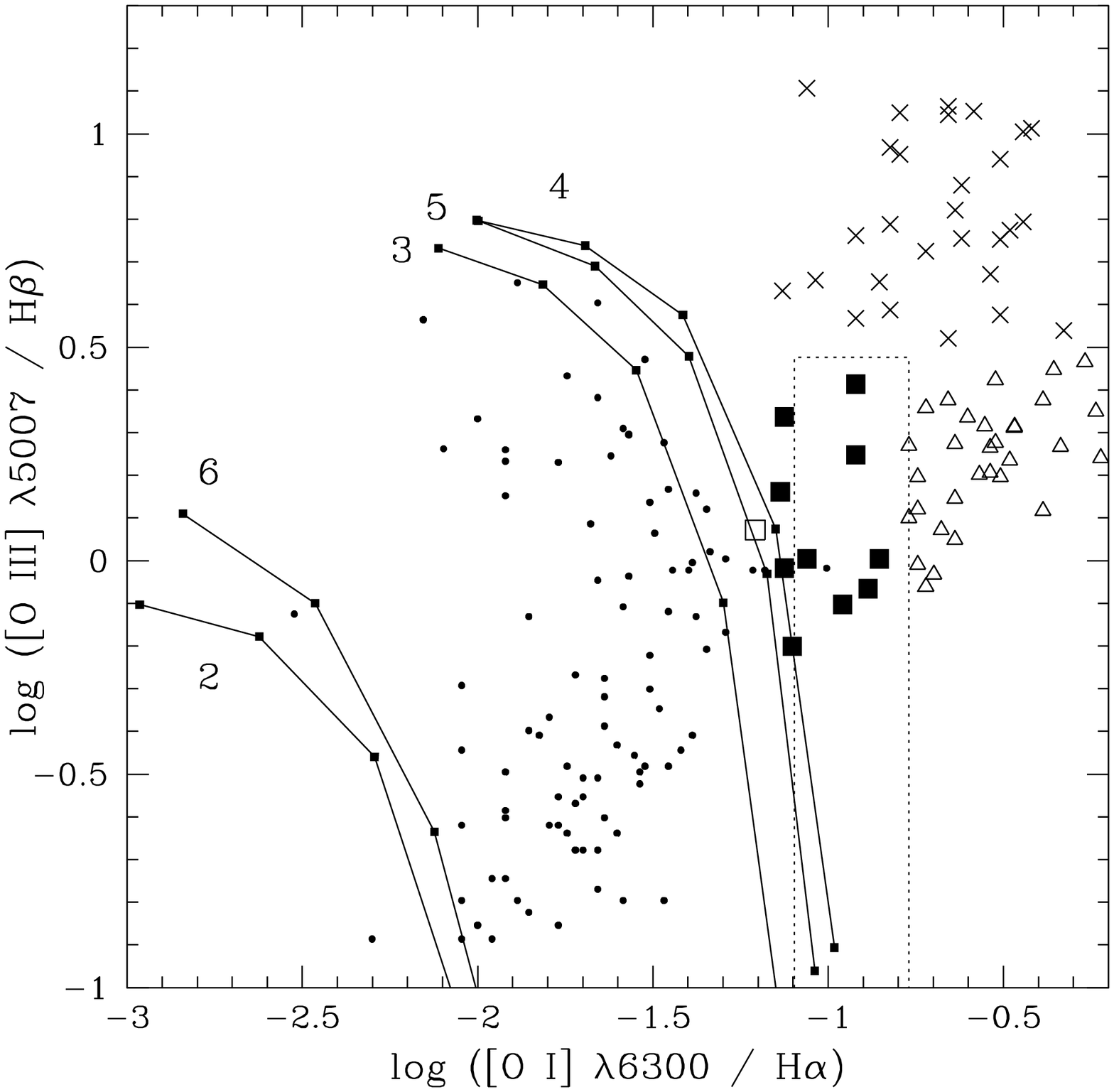]{Burst age effects on the [\ion{O}{1}] \lam6300
line strength, for model grid A.  Plot symbols are the same as for
Figure \ref{oratio}.  The model sequences are shown for nebular
density $n_e = 10^{3}$ \percucm\ and burst ages of 2, 3, 4, 5, and 6
Myr.
\label{oage}}

\figcaption[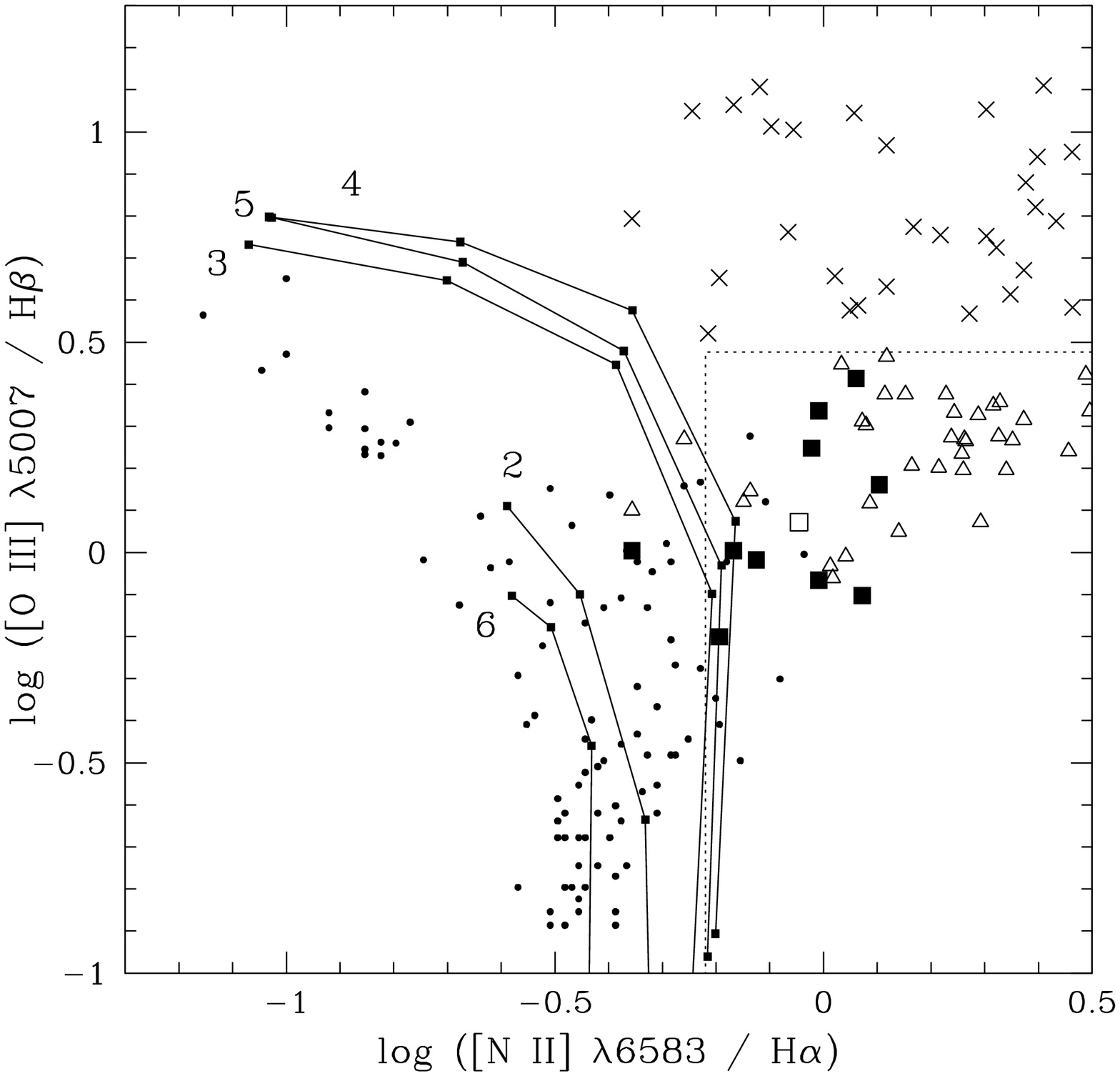]{Burst age effects on the [\ion{N}{2}] \lam6583
line strength, for model grid A. Plot symbols are the same as for
Figure \ref{nratio}.  The model sequences are shown for nebular
density $n_e = 10^{3}$ \percucm\ and burst ages of 2, 3, 4, 5, and 6
Myr.  \label{nage}}

\figcaption[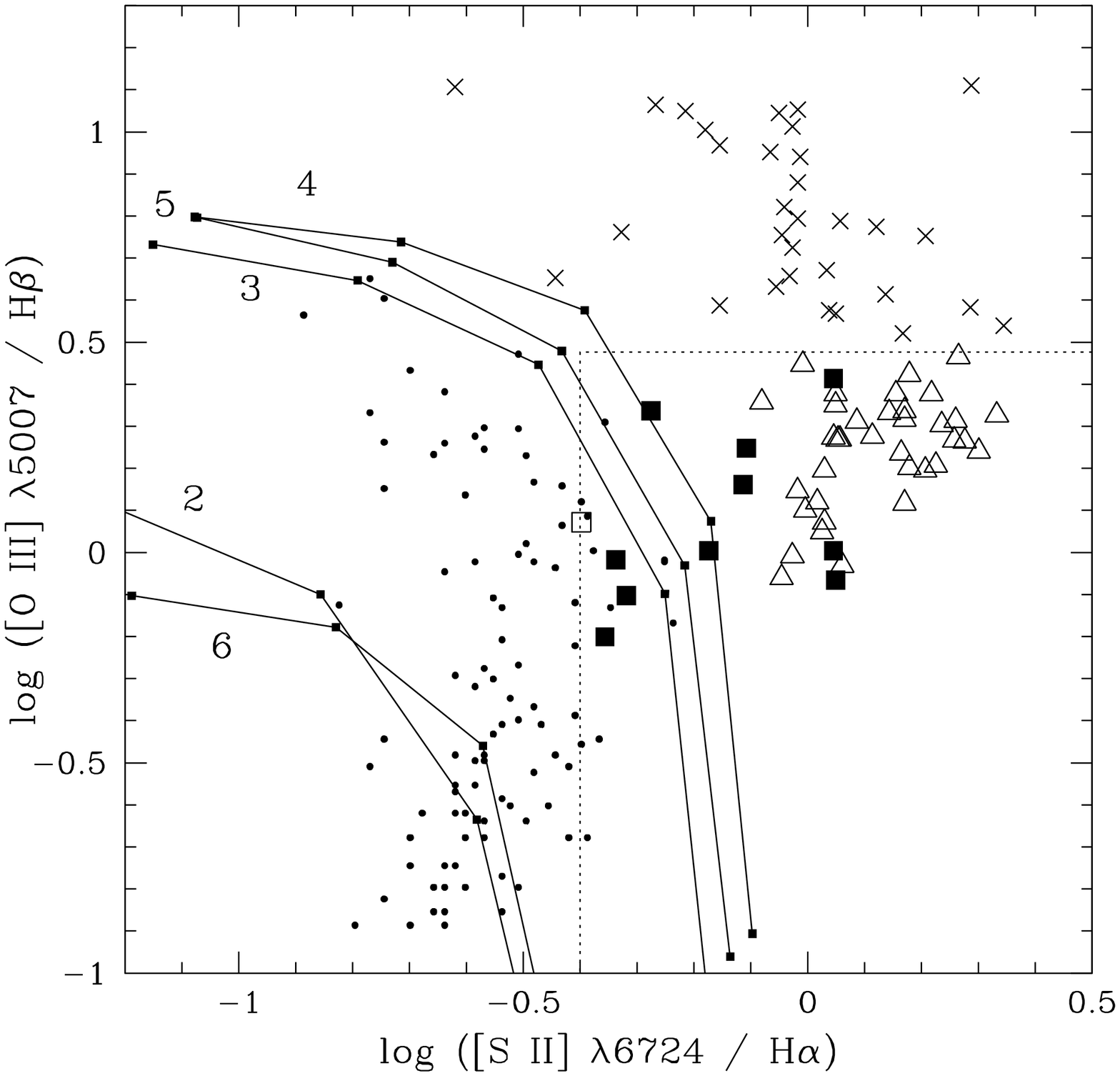]{Burst age effects on the [\ion{S}{2}]
\lamlam6716, 6731 line strength, for model grid A.  Plot symbols are
the same as for Figure \ref{sratio}.  The model are shown for nebular
density $n_e = 10^{3}$ \percucm\ and burst ages of 2, 3, 4, 5, and 6
Myr.
\label{sage}}

\figcaption[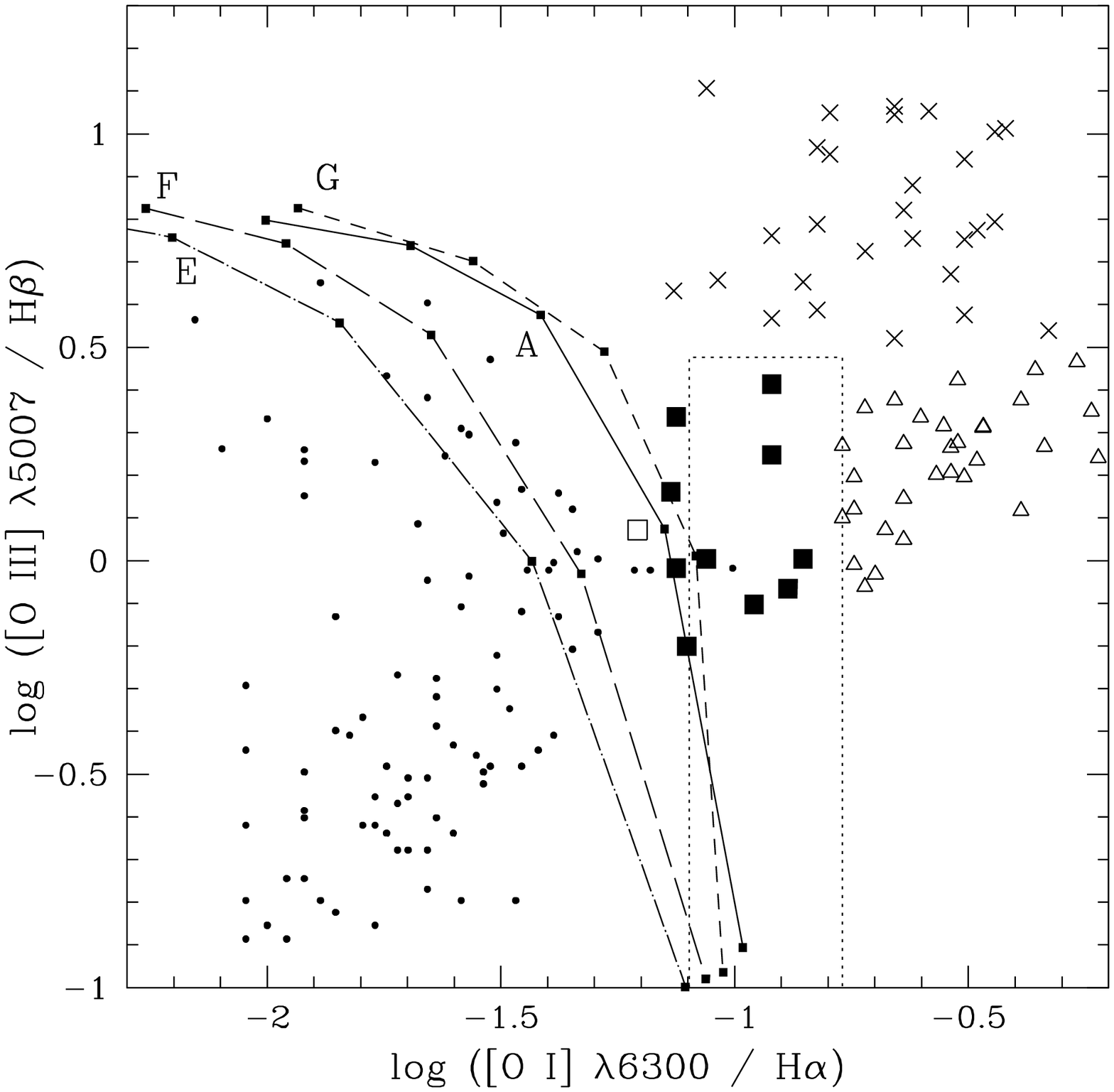]{Metallicity effects on the [\ion{O}{1}]
\lam6300 / \hal\ flux ratio.  Plot symbols are the same as for Figure
\ref{odensity}.  Model grids are plotted for the cases of $Z/\zsun$ =
0.2 (dot-dashed line; from grid E), 0.4 (long-dashed line; from grid
F), 1 (solid line; from grid A), and 2 (short-dashed line; from grid
G).  Each sequence is shown for \nh\ = $10^3$ \percucm\ and $t = 4$
Myr. \label{olometal}}

\figcaption[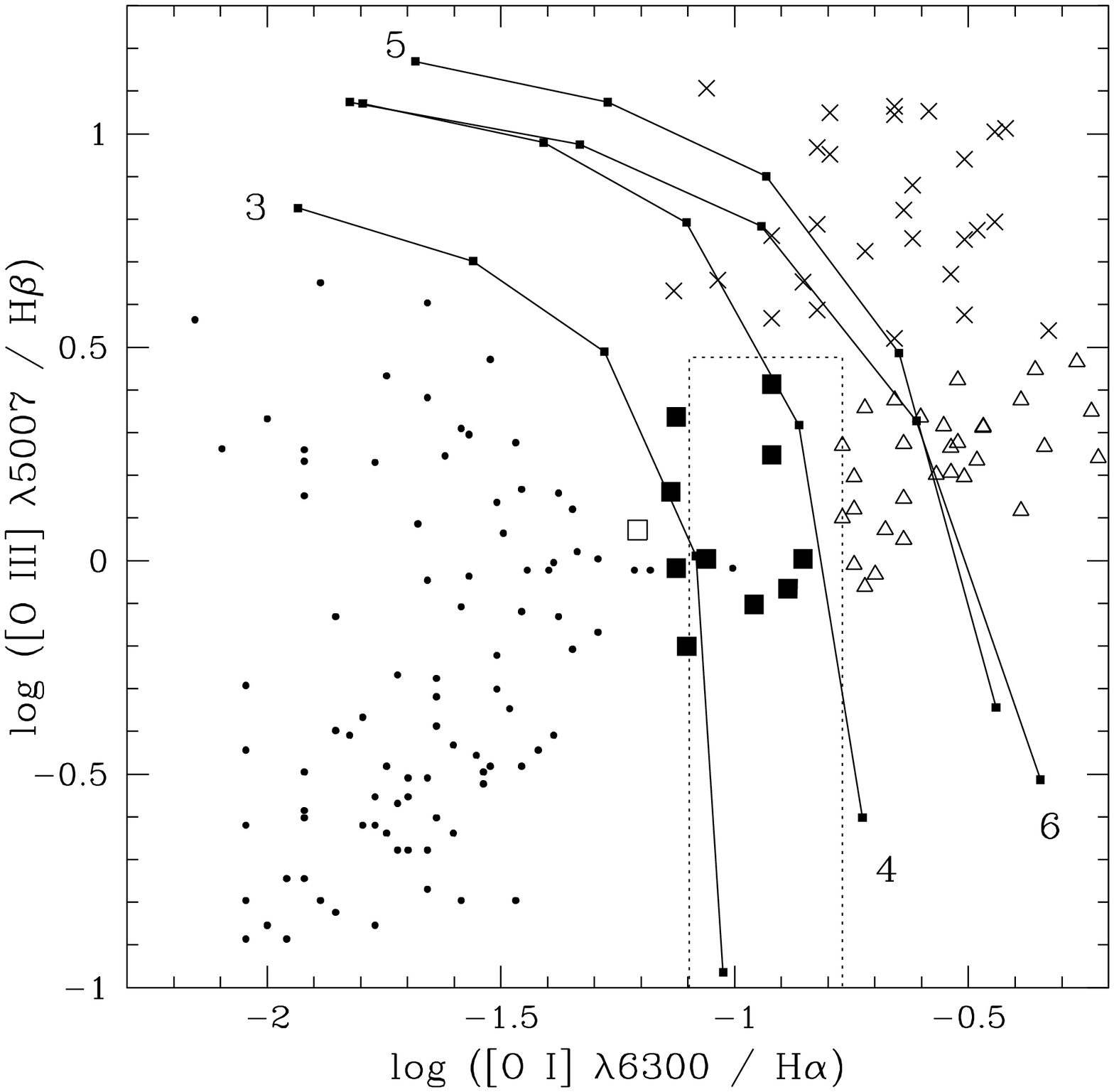]{The [\ion{O}{1}] \lam6300 / \hal\ flux ratio
for the case of $Z = 2\zsun$ (model grid G), at $t = 4$ Myr.  Plot
symbols are the same as for Figure \ref{odensity}.  The model
sequences are plotted for densities of log ($n_H$/\percucm) = 3, 4, 5,
and 6. \label{ohimetal} }

\setcounter{figure}{0}

\begin{figure}
\plotone{oratio2.ps}
\caption{}
\end{figure}

\begin{figure}
\plotone{nratio2.ps}
\caption{}
\end{figure}

\begin{figure}
\plotone{sratio2.ps}
\caption{}
\end{figure}

\begin{figure}
\plotone{odensity.ps}
\caption{}
\end{figure}

\begin{figure}
\plotone{oage.ps}
\caption{}
\end{figure}

\begin{figure}
\plotone{nage.ps}
\caption{}
\end{figure}

\begin{figure}
\plotone{sage.ps}
\caption{}
\end{figure}

\begin{figure}
\plotone{olometal.ps}
\caption{}
\end{figure}

\begin{figure}
\plotone{ohimetal.ps}
\caption{}
\end{figure}

\end{document}